\documentclass{aa}
\usepackage{natbib}
\usepackage{graphicx}
\usepackage{txfonts}
\usepackage{hyperref}
\usepackage{url}
\usepackage{xcolor}
\hypersetup{
colorlinks,
linkcolor={red!80!black},
citecolor={blue!80!black},
urlcolor={blue!80!black}
}

\def\cm3{cm$^{-3}$}
\def\kms{km~s$^{-1}$}

\def\rsun{R$_{\odot}$}

\def\msun{M$_{\odot}$}

\def\one{\ts {\,\sc i}}

\def\beq{\begin{equation}}
\def\eeq{\end{equation}}

\def\lesssim{\mathrel{\hbox{\rlap{\hbox{\lower4pt\hbox{$\sim$}}}\hbox{$<$}}}}
\def\gtrsim{\mathrel{\hbox{\rlap{\hbox{\lower4pt\hbox{$\sim$}}}\hbox{$>$}}}}

\def\one{{\,\sc i}}

\def\v1d{{\sc v1d}}
\def\kepler{{\sc kepler}}
\def\mesa{{\sc mesa}}
\def\cmfgen{{\sc cmfgen}}

\newcommand{\iso}[2]{\ensuremath{^{#1}\rm{#2}}}

\def\aj{AJ}
\def\pasp{PASP}

\def\apj{ApJ}
\def\apjs{ApJS}
\def\apjl{ApJL}
\def\aap{A\&A}
\def\araa{ARA\&A}
\def\aaps{A\&AS}
\def\mnras{MNRAS}
\def\nat{Nature}

\def\apss{Astrophysics and Space Science}

\def\nifs{\iso{56}Ni}
\def\cofs{\iso{56}Co}

\begin{document}

   \title{The difficulty of inferring progenitor masses from Type II-Plateau supernova light curves}
   \titlerunning{Type II-P SNe and their progenitors}

\author{Luc Dessart\inst{\ref{inst1}}
  \and
   D. John Hillier\inst{\ref{inst2}}
  }

\institute{
Unidad Mixta Internacional Franco-Chilena de Astronom\'ia (CNRS, UMI 3386),
    Departamento de Astronom\'ia, Universidad de Chile,
    Camino El Observatorio 1515, Las Condes, Santiago, Chile\label{inst1}
\and
    Department of Physics and Astronomy \& Pittsburgh Particle Physics,
    Astrophysics, and Cosmology Center (PITT PACC),  University of Pittsburgh,
    3941 O'Hara Street, Pittsburgh, PA 15260, USA.\label{inst2}
  }

   \date{Received; accepted}

  \abstract{
  Much controversy surrounds the inferred progenitor masses of Type II-Plateau (II-P) supernovae (SNe). The debate is nourished by the discrepant results from radiation-hydrodynamics simulations, from pre-explosion imaging, and from studies of host stellar populations. Here, we present a controlled experiment using four solar metallicity models with  zero-age main-sequence masses of 12, 15, 20, and 25\,\msun. Because of the effects of core burning and surface mass loss, these models reach core collapse as red-supergiant (RSG) stars with a similar H-rich envelope mass of 8 to 9\,\msun\ but with final masses in the range  $11$ to $16$\,\msun. We explode the progenitors using a thermal bomb, adjusting the energy deposition to yield an asymptotic ejecta kinetic energy of $1.25 \times 10^{51}$\,erg and an initial \nifs\ mass of 0.04\,\msun. The resulting SNe produce similar photometric and spectroscopic properties from 10 to 200\,d. The spectral characteristics are degenerate. The scatter in early-time color results from the range in progenitor radii, while the differences in late-time spectra reflect the larger oxygen yields in more massive progenitors. Because the progenitors have a comparable H-rich envelope mass, the photospheric phase duration is comparable for all models; the difference in He-core mass is invisible. As different main-sequence masses can produce progenitors with a similar H-rich envelope mass, light curve modeling cannot provide a robust and unique solution for the ejecta mass of Type II-P SNe. The numerous uncertainties in massive star evolution and wind mass loss also prevent a robust association with a main-sequence star mass. Light curve modeling can at best propose compatibility.
  }

   \maketitle

\section{Introduction}

   Understanding how the landscape of Type II-Plateau (II-P) supernova (SN) properties connects to the diversity of red-supergiant (RSG) star progenitors and their explosion is of great interest for astrophysics. Unfortunately discrepant estimates of the ejecta and progenitor masses are often obtained from radiation-hydrodynamics simulations that study the bolometric light curve evolution\footnote{We will refer loosely to this approach as ``light-curve modeling". In practice, it usually (but not always) includes the additional information from one spectral line to infer the evolution of the expansion rate at the photosphere.}, from pre-explosion imaging, and from studies of host stellar populations.

While significant advances have been made in the modeling of the proto-neutron star phase leading to shock revival and explosion in massive star progenitors, there are still many unresolved issues. Although the ejecta properties in these studies are broadly consistent with inferred properties from SN II-P observations \citep{lentz_ccsn_3d_15,mueller_ccsn_3d_17,glas_ccsn_3d_18,oconnor_couch_ccsn_3d_18,vartanyan_ccsn_3d_19}, the simulations cannot predict which stars produce core collapse SNe, and quantities such as the \nifs\ mass and explosion energy. Complications also arise from uncertainties in the structure of the SN progenitor prior to core collapse.  These theoretical studies would benefit from having reliable observational inferences about SN and progenitor properties.

 Using pre-explosion photometry, observations have allowed the detection of a progenitor RSG for a handful of SNe II-P \citep{vandyk_03gd_03,vandyk_08bk_12,vandyk_12aw_12,smartt_03gd_04,maund_05cs_05,maund_12ec_13,maund_08bk_14,fraser_12aw_12,fraser_13ej_14,oneill_18aoq_19}. These works propose main-sequence progenitor masses below about 17\,\msun\ (and the lack of more massive RSG progenitors has been coined the ``RSG'' problem), although these inferences are often uncertain. The problem of inferring the progenitor luminosity is made difficult, for example, because of the frequent lack of multi-band photometry, the uncertain bolometric correction \citep{davies_beasor_18}, or the uncertain reddening \citep{beasor_davies_16}. Furthermore, these inferences are based on stellar evolution models, whose predictions depend on physical properties either not known a priori (e.g., initial rotation) or hard to model (e.g, convection, overshooting; \citealt{arnett_mlt_15}). These can impact the He-core mass and the progenitor luminosity at the time of explosion \citep{heger_langer_00,meynet_maeder_00,hirschi_rot_04}.

   \begin{figure*}
   \vspace{-0.5cm}
\begin{center}
\includegraphics[width=0.48\hsize]{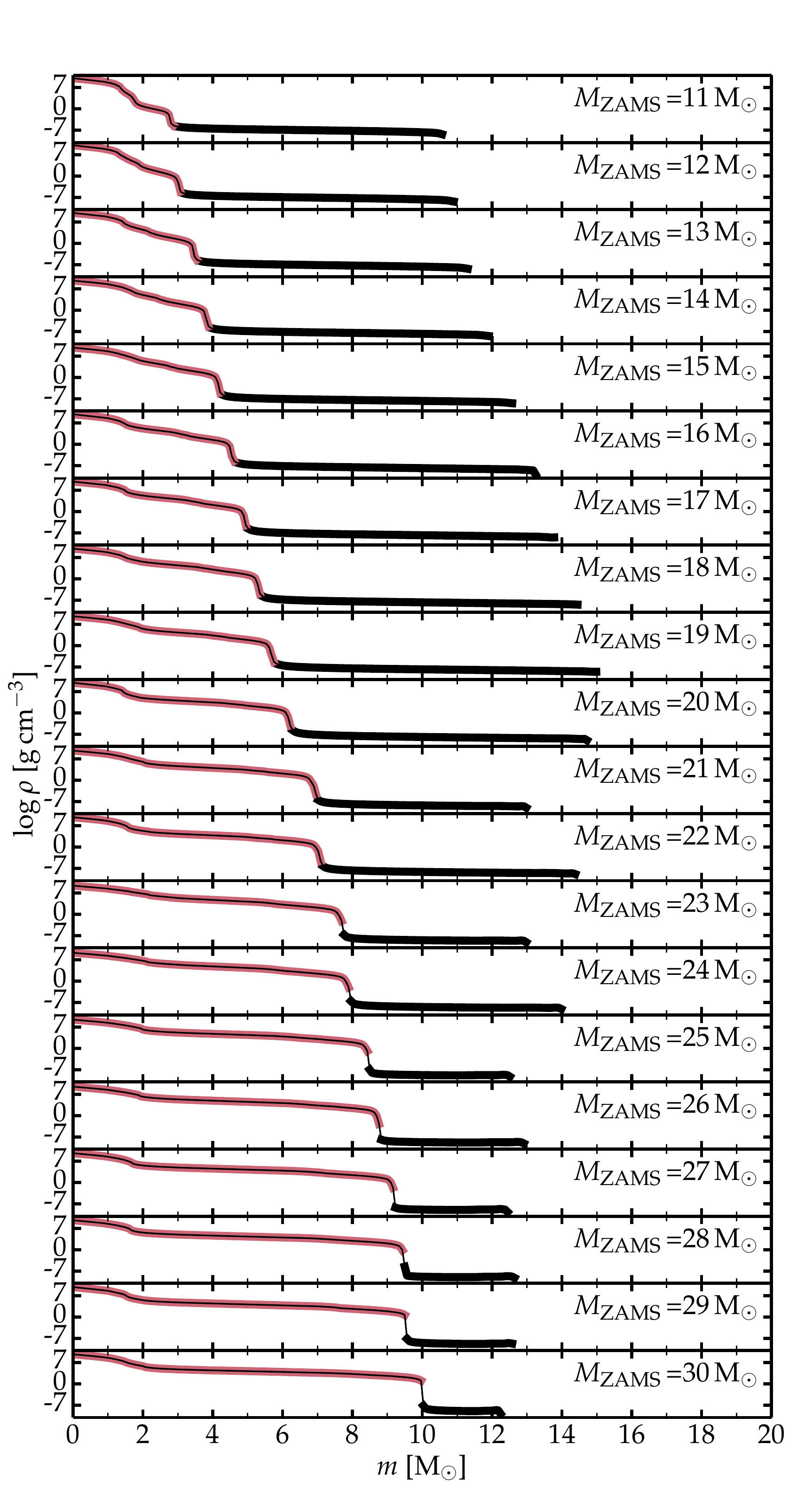}
\includegraphics[width=0.48\hsize]{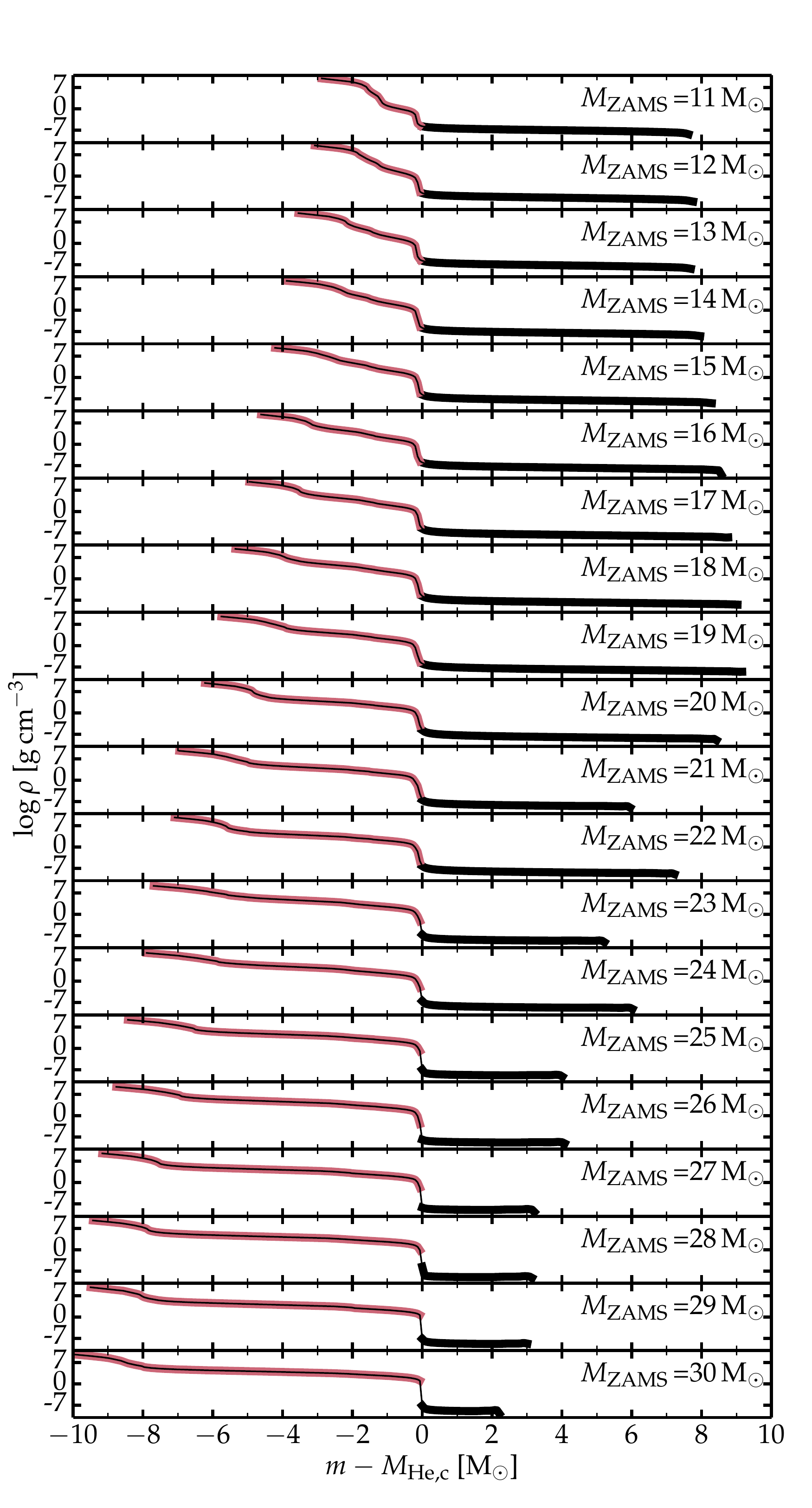}
\end{center}
   \vspace{-0.7cm}
   \caption{Density profiles versus Lagrangian mass at the time of core collapse for a set of stellar models with an initial mass on the main sequence between 11 and 30\,\msun\ \citep{whw02}. The origin is at star center (left column) or at the He core edge (right column). The H-rich envelope is drawn with a thick black line and corresponds in all models to (extended) regions with a density below about 10$^{-7}$\,g\,cm$^{-3}$. The He core is drawn with a thick red line and corresponds in all models to (compact) regions with a density above about 1\,g\,cm$^{-3}$. The vertical scale is squeezed but is suitable to reveal the evolution of the He core mass and the H-rich envelope mass for different ZAMS masses. In the left panel we see the smooth progression in the He core mass with initial progenitor mass, and the varying progenitor mass at the onset of core collapse. In the right panel the non-linear evolution of the H-rich envelope mass with initial progenitor mass is emphasized. In these models, stars with ZAMS masses of 11 to 21\,\msun\ contain similar H-rich envelope masses (8 to 9\,\msun) at the onset of core collapse.
}
   \label{fig_whw02}
\end{figure*}

   Radiation hydrodynamics and radiative transfer can also be used to characterize the properties of SNe II-P. The most robust inference and the most easily done is for the \nifs\ mass since full $\gamma$-ray trapping holds at the onset of the nebular phase. At that time, the bolometric luminosity equals the total power radiated from \nifs\ and \cofs\ decay,  and the mass of ejected \nifs\ can be determined at an accuracy that depends only on the accuracy of the adopted distance, reddening, and amount of flux falling outside of the observed range (e.g., \citealt{woosley_87A_late_88,hamuy_03}). Other inferences such as explosion energy, ejecta mass, progenitor mass at collapse or on the main sequence, require some modeling and are therefore more uncertain.

   Radiation hydrodynamic simulations of Type II-P SNe have been done for several decades. Early works identified the basic properties of the progenitors, in particular their large progenitor radii and massive H-rich envelope compatible with RSG progenitors \citep{grassberg_71,FA77}. Basic relations between ejecta and observed SN II-P properties were also drawn \citep{litvinova_sn2p_85,popov_93}. When applied to observed SNe II-P, these relations can however produce very perplexing numbers, such as ejecta masses of several 10\,\msun\ or small supergiant radii atypical of RSG stars (see, e.g., \citealt{hamuy_03}). In many cases, these quantities conflict with predictions from stellar evolution and progenitor observations.

Studies dedicated to specific objects  allow for a more refined analysis through the production of a tailored model (see, e.g., \citealt{turatto_97d_98,utrobin_07_99em,chugai_utrobin_97D_00,bersten_11_2p,d13_sn2p,lisakov_ll2p_18,morozova_sn2p_18}). But in this case, one notes a large disparity in inferred ejecta properties, in particular the ejecta mass, which then translates into an even larger disparity in the progenitor main sequence mass. These masses are often much larger than those inferred for the progenitor stars from pre-explosion imaging \citep{smartt_09}, or from the modeling of nebular-phase spectra \citep{maguire_2p_12,jerkstrand_04et_12,jerkstrand_12aw_14}. Since mass is the most defining characteristic of a star, this discrepancy is a problem.

SN II-P light curves are, however, primarily sensitive to the H-rich envelope mass and not to the mass of the progenitor He-core \citep{DH11_2p,d13_sn2p}. Hence, in many instances, the inferred ``ejecta'' mass refers only to the H-rich envelope mass of the progenitor star at the time of explosion. This ambiguity arises because the words envelope mass and ejecta mass are used inter-changeably in the literature, while the two words refer to two different entities. Stellar evolution indicates that their value may differ by a factor of ten.

To clarify this property, we utilize the public grid of massive star models from \citet{whw02}. Fig.~\ref{fig_whw02} shows a montage of density profiles at core collapse for massive stars evolved at solar metallicity with zero-age main sequence masses ($M_{\rm ZAMS}$) of 11 up to 30\,\msun. The left column shows the standard way of presenting such profiles, with the origin at the center of the star. As the initial mass $M_{\mathrm ZAMS}$ increases, the mass of the He core ($M_{\rm He,c}$; shown in red, and corresponding to regions with a density greater than about 1\,g\,cm$^{-3}$) increases from about 3\,\msun\ for the 11\,\msun\ model up to 10\,\msun\ for the 30\,\msun\ model. It is in this hot and dense He core that nuclear fusion takes place. In the lower-density lower-temperature H-rich envelope, the composition deviates from its primordial value only through the effect of mixing (see, e.g., \citealt{davies_dessart_19}). The mass of the H-rich envelope ($M_{\rm H,e}$) at core collapse is between 8 and 9\,\msun\ for stars with  $M_{\mathrm ZAMS}$ between 11 and 21\,\msun, and decreases (not necessarily monotonically) as  $M_{\mathrm ZAMS}$ increases further. This arises from an empirically-based formulation of RSG mass loss rates, which reflects the greater mass loss rates inferred for more massive and more luminous RSGs \citep{dejager_mdot_88}. This influence of mass loss on $M_{\rm H,e}$ is what eventually turns more massive stars into H-deficient Wolf-Rayet stars \citep{maeder_meynet_87,langer_wr_94,pac_wr_07}. While RSG star mass loss remains uncertain, this should only affect the value of  $M_{\mathrm ZAMS}$ where $M_{\rm H,e}$ starts going down with increasing $M_{\mathrm ZAMS}$, and the actual value of $M_{\rm H,e}$ for a given  $M_{\mathrm ZAMS}$.\footnote{These quantities should also vary with metallicity.} The trend shown in Fig.~\ref{fig_whw02} must hold.

Because SN II-P light curves are primarily sensitive to the H-rich envelope mass, it is more instructive to show the density structure with respect to the edge of the He core (right column of Fig.~\ref{fig_whw02}). Ignoring the He core, the models with $M_{\mathrm ZAMS}$ below 21\,\msun\ have similar density structures, while above the models have progressively lower values of  $M_{\rm H,e}$. For a given ejecta kinetic energy $E_{\rm kin}$ imparted to the H-rich envelope, models with $M_{\mathrm ZAMS}$ below 21\,\msun\ should have comparable light curves (modulo differences in $R_\star$), while above 21\,\msun, the light curve should be characterized by a faster post-breakout luminosity decline and a shorter photospheric phase \citep{bartunov_blinnikov_92}. Furthermore, "ejecta" masses (i.e., $M_{\rm H,e}$ values) inferred from SN II-P light curves should consequently be around 10\,\msun\ since the RSG stars from $10-20$\,\msun\ stars on the main sequence are favored by the initial mass function. Masses of just a few \,\msun\ (which result from higher mass progenitors) should be less frequently inferred (the reasoning here ignores binarity). This basic picture seems compatible with the observed diversity of Type II SN light curves \citep{anderson_2pl,sanders_sn2_15}. And the large range of Type II SN ejecta masses (extending to large masses) published in the literature appears incongruous.

\begin{table}
\caption{Properties of the model set, including the final mass at core collapse ($M_{\rm fin}$), the He-core mass ($M_{\rm He,c}$), the H-rich envelope mass ($M_{\rm H,e}$), the surface radius ($R_\star$), the ejecta kinetic energy ($E_{\rm kin}$), the ejecta mass ($M_{\rm ej}$), the oxygen mass ($M({\rm O})$), and the \nifs\ mass. The initial mass on the main sequence is reflected in the model name, i.e., 12\,\msun\ for model m12. Numbers in parenthesis represent powers of ten.
\label{tab_sum}
}
\begin{center}
\begin{tabular}{l@{\hspace{1.5mm}}c@{\hspace{1.5mm}}c@{\hspace{1.5mm}}
c@{\hspace{1.5mm}}c@{\hspace{1.5mm}}c@{\hspace{1.5mm}}
c@{\hspace{1.5mm}}c@{\hspace{1.5mm}}c@{\hspace{1.5mm}}
}
\hline
  & $M_{\rm fin}$ & $M_{\rm He,c}$ & $M_{\rm H,e}$  & $R_\star$ & $E_{\rm kin}$ & $M_{\rm ej}$ & $M({\rm O})$ & $M$(\nifs) \\
     & [\msun] & [\msun] & [\msun] & [\rsun] & [erg] & [\msun] & [\msun]  & [\msun]  \\
\hline
 m12    & 11.21     &    3.26  & 7.95    &  406   &     1.27(51) &  9.79   & 0.16 & 0.046  \\
 m15   & 13.12   &     4.56  & 8.56   &   589   &     1.28(51) & 11.57  & 0.53  & 0.041   \\
 m20    & 14.86   &     6.81  & 8.05   &   843   &    1.25(51)  & 13.18 & 1.33  & 0.041   \\
 m25   & 15.96   &     8.59   & 7.37  &   872   &    1.21(51)  & 13.93  & 2.33  & 0.042 \\
\hline
\end{tabular}
\end{center}
\flushleft
\end{table}

\begin{figure}[t]
\includegraphics[width=\hsize]{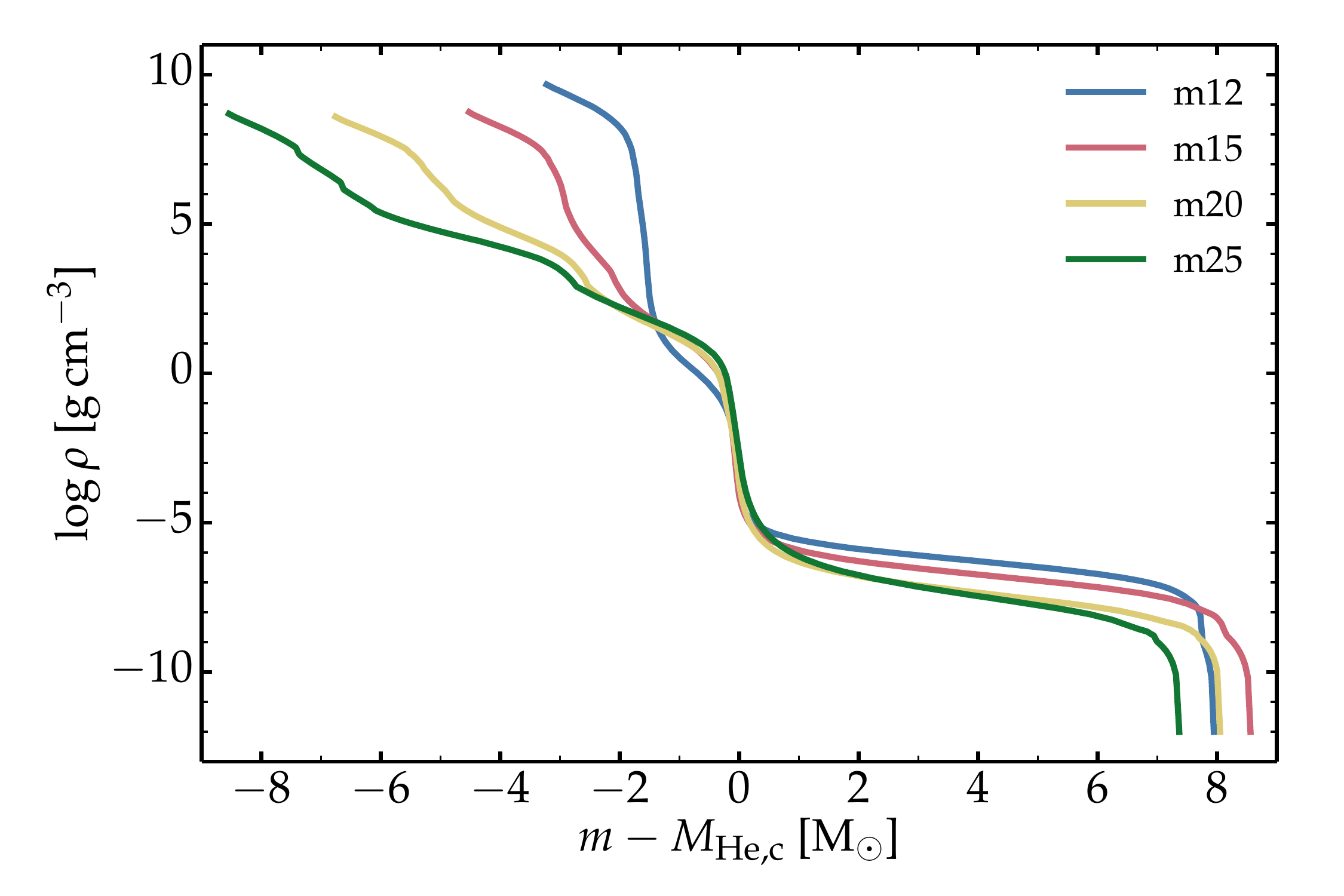}
   \caption{Density structure of the models m12, m15, m20, and m25 shown versus Lagrangian mass (the origin is the outer edge of the He core). The differences in H-rich envelope density reflect the differences in surface radii (see Table~\ref{tab_sum}).}
   \label{fig_mesa}
\end{figure}
 \begin{figure}[t]
\includegraphics[width=\hsize]{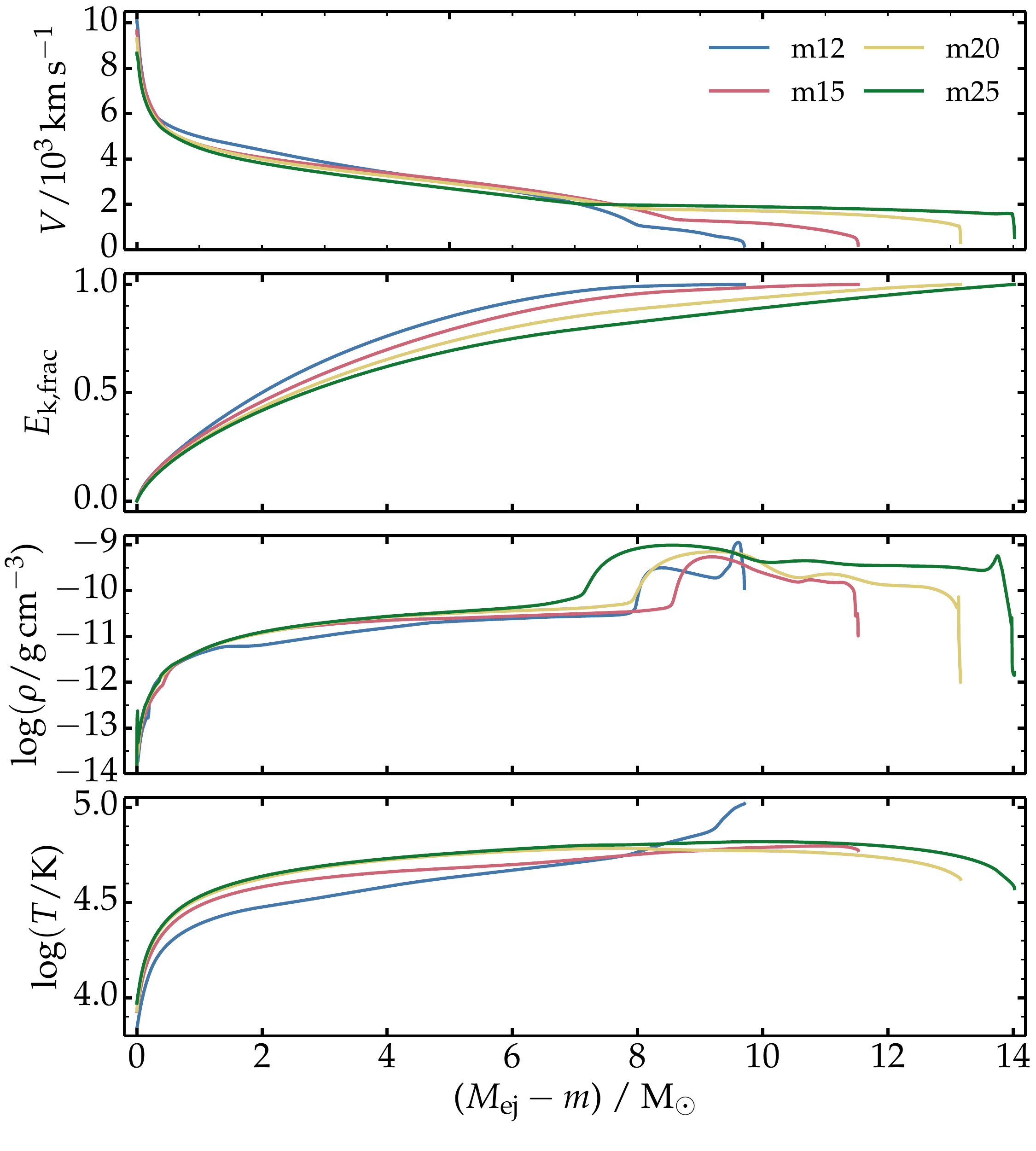}
   \caption{Ejecta velocity, fractional kinetic energy $E_{\rm k,frac}$, density, and temperature versus mass at 14\,d after the piston trigger. $E_{\rm k,frac}$ is defined as $\int V^2 dm / 2E_{\rm kin}$, with the integration done inward. The x-axis origin is at the outermost ejecta layer, to emphasize how similar the properties of the shocked H-rich envelope are in these different progenitor models.
   \label{fig_v1d}
   }
   \end{figure}

The goal of this paper is to demonstrate that the SN II-P ejecta masses inferred from light curve modeling bear considerable uncertainty. One cannot infer the progenitor mass (either at core collapse or on the main sequence) using light curve modeling. The problem is degeneracy since stars with widely different total masses may have the same $M_{\rm H,e}$ at core collapse. To illustrate this property, we conduct a controlled experiment on a set of massive star models with different main-sequence mass but the same H-rich envelope mass -- our model set is very similar to that of \citet{whw02} and thus representative of predictions for single star evolution. The next section presents the numerical approach. Section~\ref{sect_res} then presents the SN radiative properties for each model and discusses the implications of the results.

\section{Numerical approach}
\label{sect_setup}

The progenitor models correspond to solar-metallicity non-rotating stars of 12, 15, 20, and 25\,\msun\ on the main sequence. These were evolved with \mesa\ version 7623 \citep{mesa1,mesa2,mesa3} as part of earlier works \citep{lisakov_08bk_17,lisakov_ll2p_18}. These stars are evolved as single stars, or in binary systems with a wide orbit: the possibility of mass exchange with a companion star is ignored. These simulations use standard \mesa\ default parameters, except for a mixing length parameter of three. This choice is needed to produce more compact RSG stars at core collapse (see \citealt{d13_sn2p}, and more recently \citealt{mesa4}). Because this produces RSG star models with a higher effective temperature, a slight enhancement on the default mass-loss rate parameter is needed in order to yield a mass loss rate similar to that of RSGs with a lower effective temperature.  The model properties are summarized in Table~\ref{tab_sum}, while Fig.~\ref{fig_mesa} shows the density structure at the onset of core collapse for each model. The models differ in He-core mass but have a similar H-rich envelope mass -- their properties are similar to those of \citet{whw02} for the corresponding ZAMS mass (Fig.~\ref{fig_whw02}). In radial space, the differences in He core properties are invisible since the He core has the size of the Sun while the surface radii cover approximately from 400 to 900\,\rsun.  Because they make no distinction between the high-density He core and the low density H-rich envelope, polytropic progenitor structures are unsuited for light curve modeling. To comply with the early-time observations of some SNe II-P (see, e.g., \citealt{yaron_13fs_17}), about $\sim$\,0.2\,\msun\ of atmospheric (static) material is added above $R_\star$ using a density scale height of about 0.2\,$R_\star$. This has however no relevance for the present discussion.

\begin{figure}[t]
\includegraphics[width=\hsize]{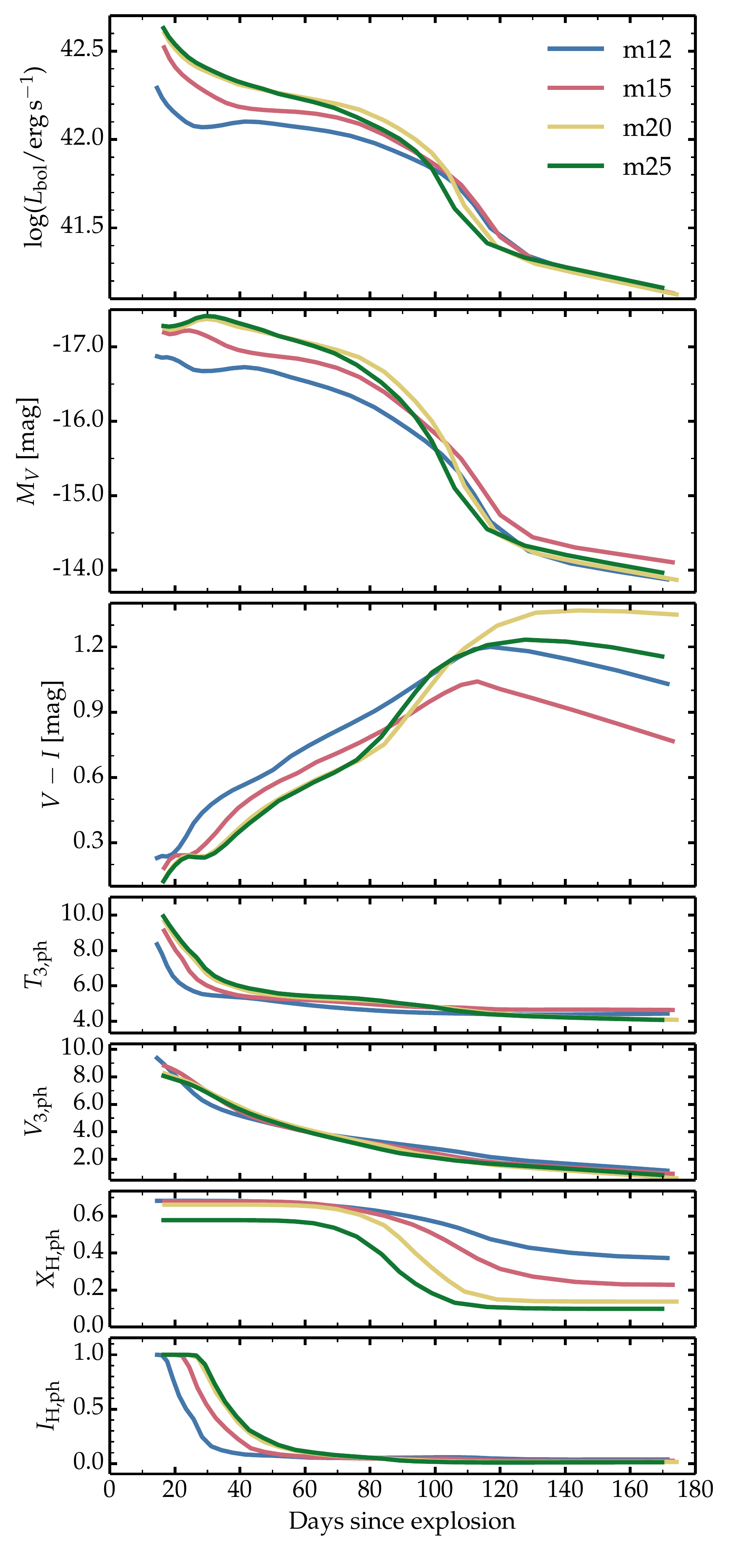}
 \vspace{-0.9cm}
  \caption{Summary of results from the \cmfgen\ simulations of models m12, m15, m20, and m25, showing, from top to bottom, the bolometric light curve, the $V$-band light curve, the $V-I$ color evolution, and the evolution of some photospheric properties ($T_{\rm 3, ph}$: temperature in units of 1000\,K; $V_{\rm 3, ph}$: velocity in units of 1000\,\kms; $X_{\rm H, ph}$: hydrogen mass fraction;  $I_{\rm H, ph}$: hydrogen ionization state).}
   \label{fig_all}
\end{figure}

The explosion is modeled with the 1D gray radiation-hydrodynamics code \v1d\ in the usual manner \citep{livne_93,dlw10b}. The explosion is triggered by means of a piston placed within the Si-rich shell, at a mass cut of 1.5 to 1.94\,\msun. To compensate for the progenitor binding energy, the explosion energy is adjusted for each model to yield an asymptotic ejecta kinetic energy $E_{\rm kin}$ of $\sim 1.25 \times 10^{51}$\,erg. Although \v1d\ treats explosive nucleosynthesis, the \nifs\ mass is reset to be 0.04\,\msun\ at 100\,s after the piston trigger (the value in Table~\ref{tab_sum} deviates a little because of the remapping into \cmfgen). At 14\,d after the piston trigger (when we remap into \cmfgen), the four models have essentially the same density structure in velocity space all the way down to the He-core material (whose outer location corresponds to the density jump; Fig.~\ref{fig_v1d}).  The jump is at a velocity of $1000-2000$\,\kms\ greater in model m25 than in model m12, which reflects the differences in He-core mass between the models (see \citealt{dlw10b} for discussion). The temperature above that jump is lower in more compact progenitor models. The material that used to be in the He core has cooled considerably because it expanded by a factor of about 1000, while the H-rich envelope has expanded by a factor of about 10. This conspires to make the temperature very uniform throughout most of the ejecta. The relatively small volume occupied by the former He core stores a modest radiative energy. Instead, the large volume occupied by the shocked H-rich envelope stores most of the radiative energy that will be released during the high-brightness phase of a SN II-P. It also stores the bulk of the kinetic energy.

At 14\,d, the ejecta models computed by \v1d\ are remapped into the non local thermodynamic equilibrium and time dependent radiative transfer code \cmfgen\ \citep{HD12}. The evolution is then followed until 300\,d using the standard procedure (see, e.g., \citealt{d13_sn2p}).

There is at present no robust theory that predicts with certainty the ejecta kinetic energy and \nifs\ mass that a given ZAMS mass should produce. However, the simulations of \citet{sukhbold_ccsn_16} suggest that stars in a broad mass range from about 12 to 25\,\msun\ can yield an ejecta within 50\% of $1.2 \times 10^{51}$\,erg and a \nifs\ mass within a factor of two of 0.04\,\msun\ (their is some scatter and some offset with $M_{\mathrm ZAMS}$ depending on the calibration used for their ``thermal bomb"). Our choice is thus not unrealistic.

\section{Results and discussion}
\label{sect_res}

Figure.~\ref{fig_all} summarizes the main results from the \cmfgen\ simulations for models m12, m15, m20, and m25. The bolometric light curves are similar. Models with larger progenitor radii are more luminous, as expected (see., e.g., \citealt{popov_93,KW09,d13_sn2p}). The $V$-band light curve and the $V-I$ color evolution are also similar, with the bigger progenitors reddening later and being visually brighter (this is the same as for models s15 and s25 in \citealt{DH11_2p} and for models m15mlt1 and m15mlt3 in \citealt{d13_sn2p}). These color offsets are understood from the offset in photospheric temperature (also visible in the H ionization state), with the bigger progenitors remaining hotter at the photosphere for longer. Despite the large differences in progenitor and ejecta masses, the four models have a similar photospheric-phase duration because they have a similar H-rich envelope mass (Fig.~\ref{fig_mesa}). The models show the same qualitative and quantitative evolution in photospheric velocity, because the bulk of the kinetic energy is held up in what used to be the H-rich envelope.  Throughout the photospheric phase, the H mass fraction at the photosphere is comparable in all models (i.e., the photosphere resides in the former H-rich envelope; \citealt{DH11_2p}) which implies that the spectra  should reveal lines from similar elements.

While the light curves and colors show some differences, these are not sufficient to unambiguously determine the ejecta mass. The main driver behind the differences in photometric properties is the progenitor radius, not the ejecta mass. If we had used a different mixing length parameter for the four progenitor models (smaller in lower mass models), we could have produced progenitors with the same radius and in that case, the light curves would have been even more similar. Furthermore, because RSG observations probe the outer low-density fluff of the RSG atmosphere rather than the higher density hydrostatic photosphere (as defined in stellar evolution models), surface radii are poorly constrained. Further, and contrary to a generally-held belief, RSG radii are not predicted with any confidence by stellar evolution models since $R_\star$ is controlled by convection and a prescribed mixing length \citep{maeder_meynet_87,d13_sn2p}.

Other factors also influence the light curves.  For example, there is increasing evidence that the early-time light curves of many Type II-P SNe are influenced by circumstellar (CSM) interaction, which affects both the brightness and colors of the SN at early times \citep[e.g.,][]{yaron_13fs_17,morozova_2l_2p_17,morozova_sn2p_18,d18_13fs,moriya_13fs_17,forster_csm_18}. Clumping may also hasten the photosphere recession and boost the SN luminosity during the high-brightness phase \citep{d18_fcl}. Further, most simulations of core-collapse SNe lead to asymmetric explosions. This may affect both the light curve evolution and colors. Observations of Type II-P SNe also show variable levels of polarization, and the degree of polarization is time-dependent \citep{leonard_04dj_06,leonard_12aw_12}. The latter may indicate (although not necessarily) that the degree of asymmetry varies with time \citep{DH11_pol}. For example, the core may be much more asymmetric than the massive H-rich envelope. The amount of \nifs\ mixing is also known to affect the shape of the light curve \citep[see, e.g.,][]{KW09}, and there are observational issues related to the distance of the SN,  reddening and the inference of the bolometric light curve from a finite number of bands. Finally, we stress that even if one can infer the ejecta mass from the light curve, an extrapolation to the progenitor mass is fraught with biases because RSG mass-loss rates, and hence the progenitor mass-loss history, are poorly known (see, e.g., Fig.~1 in \citealt{meynet_rsg_15}).

\begin{figure}[t]
\begin{center}
\includegraphics[width=\hsize]{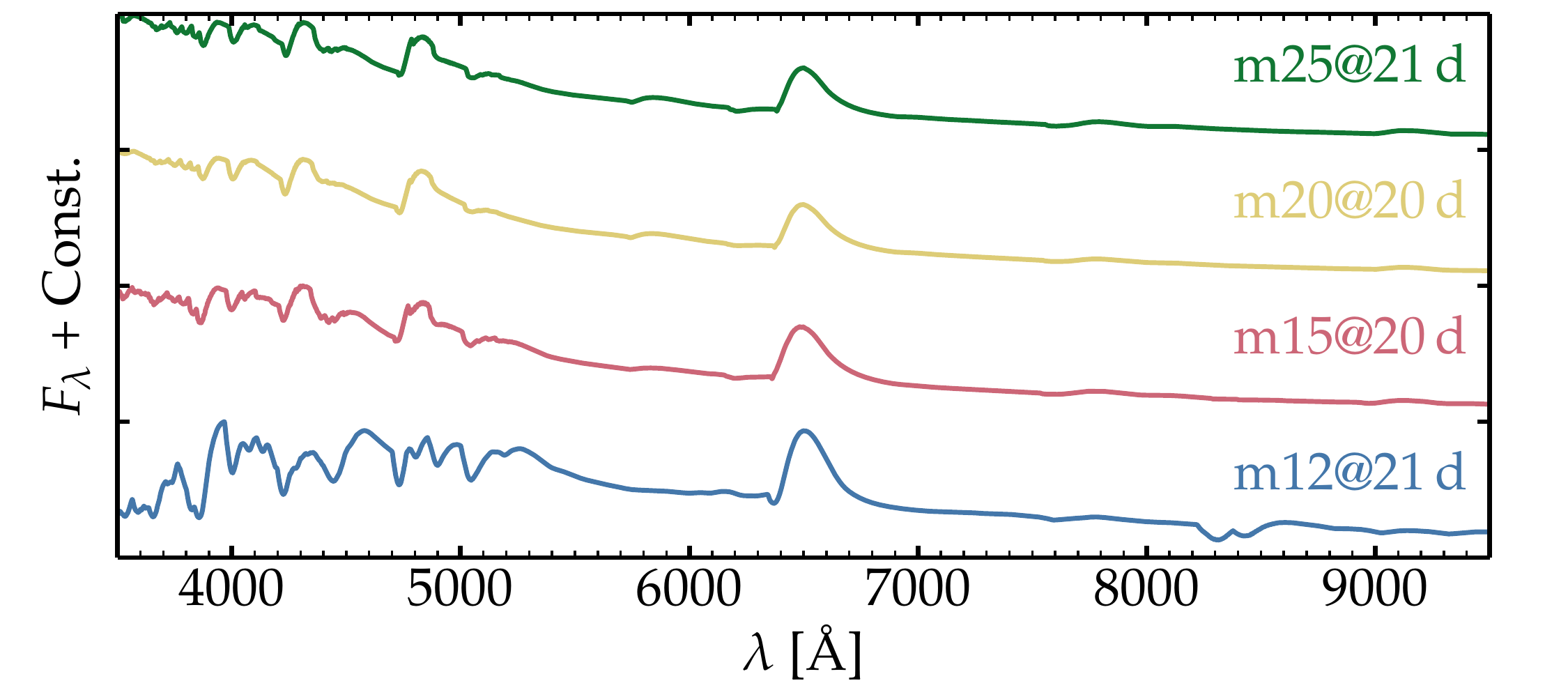}
\includegraphics[width=\hsize]{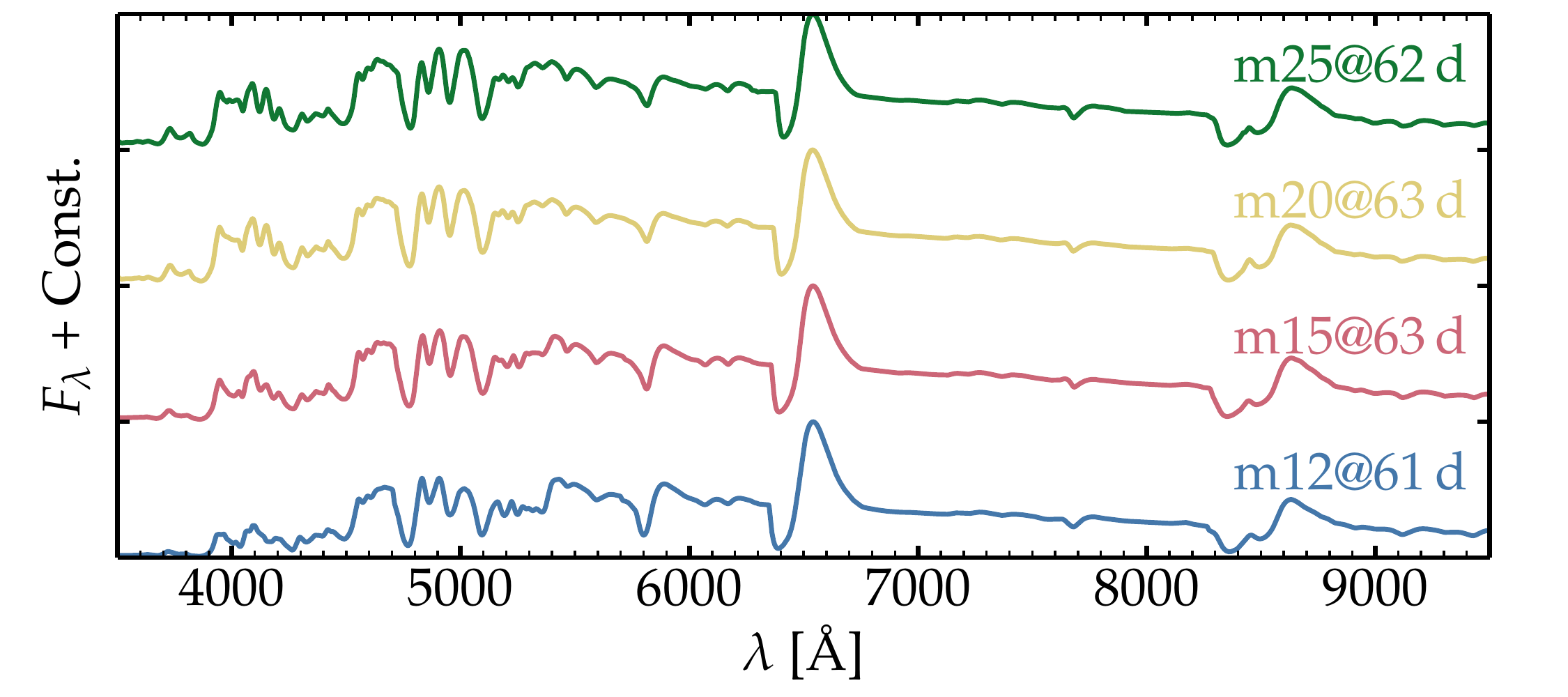}
\includegraphics[width=\hsize]{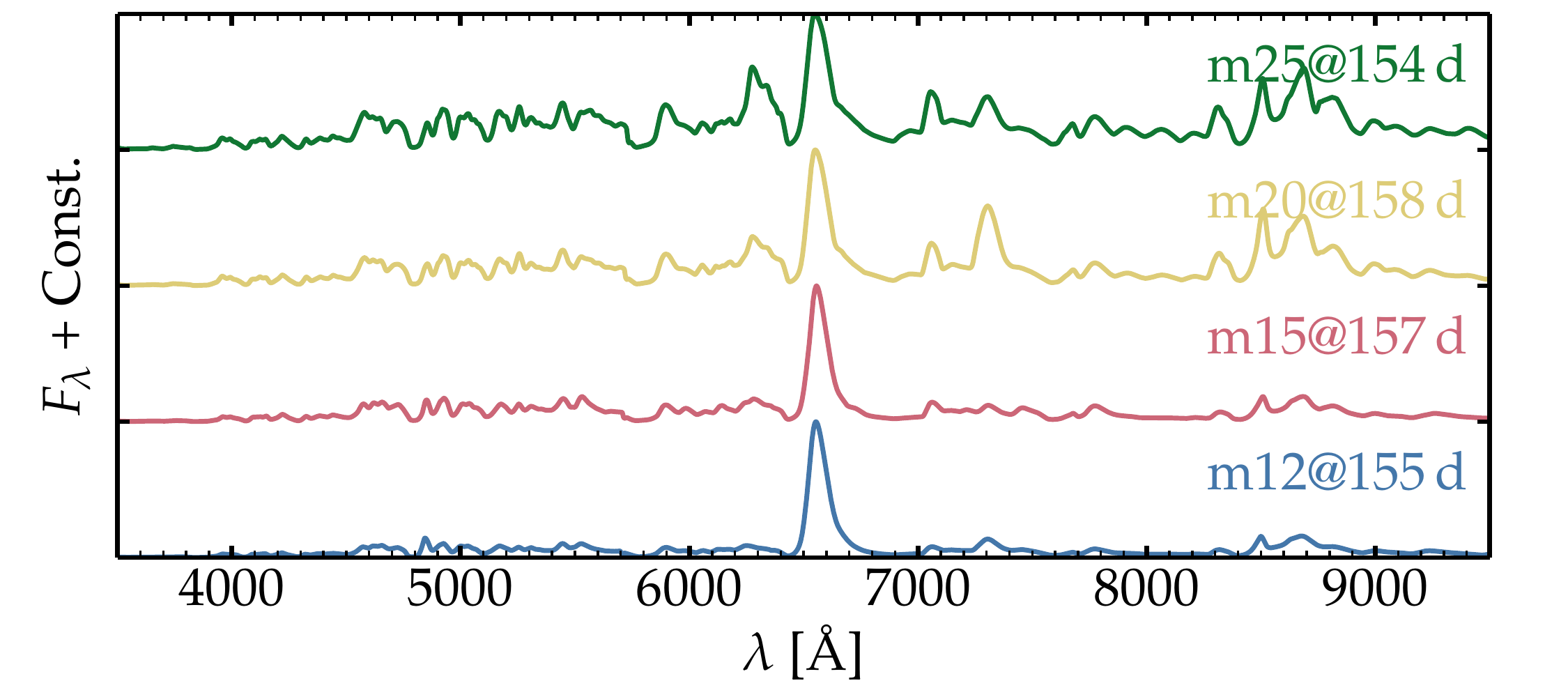}
\end{center}
 \vspace{-0.2cm}
  \caption{Comparison of optical spectra for models m12, m15, m20, and m25 at about 20\,d (top), 62\,d (middle), and 155\,d (bottom) after explosion. The color offset at early times results from the different progenitor radii. The greater O\one\,6300\,\AA\ line in more massive progenitors arises from their greater oxygen content.}
   \label{fig_spec}
\end{figure}

In Nature, it is possible that the explosion of progenitors with very different ZAMS masses yield the same asymptotic ejecta kinetic energy (which is what we assumed here for our model set) and therefore naturally contribute to light curve degeneracy. The power source for the explosion is gravitational contraction of the core and mass accretion onto it (see, e.g., \citealt{burrows_goshy_93}). This process may release more energy in higher mass progenitors, but this additional energy would be sapped by the greater binding energy of their He core. In the present set of models, to produce an ejecta kinetic energy of $1.25 \times 10^{51}$\,erg, the total energy released by the ``thermal bomb" had to be larger in more massive progenitors. Specifically, it was 1.36 (m12), 1.39 (m15), 1.78 (m20), and $1.96 \times 10^{51}$\,erg (m25) --- these values depend on the time left before the onset of collapse (e.g., the central density is not the same in all models). The trend nonetheless holds (see, e.g.,\citealt{whw02}).

Polytropes and non-evolutionary progenitor structures, designed with considerable freedom, conflict with the fundamental features of massive stars at death (primarily through the incorrect treatment of their core-halo structure). It is from these simulations that the largest discrepancies in progenitor masses arise (see also Section~4.1 in \citealt{morozova_sn2p_18}). Similarly, analytic scalings (e.g., \citealt{litvinova_sn2p_85}) are unable to yield a reliable ejecta mass since they are insensitive to the He core mass.

An extensive discussion about progenitor masses is provided by \citet{dallora_12aw_14}. In order to examine the parameter space they considered both a semi-analytic approach, and an approach based on direct hydrodynamical modeling. Their semi-analytic approach yields multiple minima in $\chi^2$ when comparing model predictions with observations. While not all gave consistency with the inferred photospheric velocities, it does highlight degeneracies in inferring ejecta masses from light curves.

\cite{morozova_sn2p_18} have also performed  a parameter study of Type II-P light curves. Unlike the study of \cite{dallora_12aw_14}, their simulations are based on evolutionary models computed with \kepler. With some exceptions, the 2D-$\chi^2$ plots of \cite{morozova_sn2p_18}  generally show a broad range of ZAMS masses and radii that are consistent with their light curves. Given the degeneracies, assumptions in both the evolutionary models and in the light curve modeling will significantly influence the results. Furthermore, \cite{morozova_sn2p_18} infer explosion energies and ejecta masses without using any spectral information, hence ignoring fundamental constraints on the expansion rate. For example, their ejecta kinetic energies are lower than standard, and with their adopted CSM mass, their ejecta models most likely fail to produce the broad Doppler-broadened lines seen in Type II SN spectra. Disregarding spectral constraints, light curve modeling is even more subject to degeneracies.

While more difficult to model, spectra potentially provide another avenue to constrain ejecta and progenitor mass. Figure~\ref{fig_spec} shows the spectra for our model set at about 20, 62, and 155\,d after explosion. At the first epoch, the difference is the greatest and results from the larger radii in more massive progenitors (which cause the brightness and color offset shown in Fig.~\ref{fig_all}). However, during the recombination phase (second epoch), the spectra are essentially identical. At the nebular epoch, the spectra are also very similar. They exhibit a strong H$\alpha$ line, but with a stronger O\one\,\,6300\,\AA\ doublet line in more massive progenitors.

A more reliable discriminant for progenitor mass may thus be sought from nebular-phase spectra. At such times, the stark contrast between a 12 and a 25\,\msun\ progenitor can be revealed from the inspection of emission lines, and in particular O\,\one\,6300\,\AA. For the present models m12 to m25, the oxygen mass increases with the ZAMS mass (it is 0.16, 0.53, 1.33, and 2.33\,\msun), and this conspires to produce a greater O\,\one\ line strength in more massive progenitors, which is apparent here in the bottom panel of Fig.~\ref{fig_spec}. Careful modeling is however needed to infer an accurate progenitor mass \citep{maguire_2p_12,jerkstrand_04et_12,jerkstrand_12aw_14}.

Our results show that models from a 12, 15, 20, and 25\,\msun\ star on the main sequence can yield a SN II-P with a similar plateau duration and spectral evolution if they have the same H-rich envelope mass and ejecta kinetic energy. Differences arise then primarily from offsets in progenitor radius (affecting the early-time brightness and color) and nucleosynthetic yields (affecting the lines from intermediate mass elements at nebular times). Because of the unconstrained mass contribution from the He core, light curve modeling cannot yield a robust inference of the ejecta mass, nor of the progenitor mass at the time of explosion. And stars with different main-sequence mass can die with the same $M_{\rm H,e}$ (which is function of the uncertain cumulative mass loss), so that the constraint of $M_{\rm H,e}$ from light curve modeling cannot be uniquely connected to a $M_{\mathrm ZAMS}$. Overall, there may well be a ``RSG problem" but light curve modeling will not solve it.

One may argue that this position is pessimistic. For example, from their set of progenitor stars and explosion models, \citet{sukhbold_ccsn_16} obtain a trend in asymptotic ejecta $E_{\rm kin}$ and \nifs\ mass, and we may eventually have a robust prediction of these quantities from ab-initio three-dimensional simulations of the explosion mechanism. We may also eventually have a robust theory for convection, overshoot, and RSG mass loss, with an accurate way to constrain core rotation. Similarly, we may eventually be able to perform multi-dimensional radiative transfer in non-LTE, thereby accounting accurately for the treatment of ejecta asymmetry, clumping, chemical segregation and mixing. When this time comes, we will be able to constrain ejecta and progenitor masses with great confidence. We will have great accuracy in our results so the degeneracies discussed in this paper will be lifted. However, this is not the present situation and even the state of the art remains limited by our understanding of each of the above building blocks.


\begin{thebibliography}{70}
\expandafter\ifx\csname natexlab\endcsname\relax\def\natexlab#1{#1}\fi

\bibitem[{{Anderson} {et~al.}(2014){Anderson}, {Gonz{\'a}lez-Gait{\'a}n},
  {Hamuy}, {Guti{\'e}rrez}, {Stritzinger}, {Olivares E.}, {Phillips},
  {Schulze}, {Antezana}, {Bolt}, {Campillay}, {Castell{\'o}n}, {Contreras}, {de
  Jaeger}, {Folatelli}, {F{\"o}rster}, {Freedman}, {Gonz{\'a}lez}, {Hsiao},
  {Krzemi{\'n}ski}, {Krisciunas}, {Maza}, {McCarthy}, {Morrell}, {Persson},
  {Roth}, {Salgado}, {Suntzeff}, \& {Thomas-Osip}}]{anderson_2pl}
{Anderson}, J.~P., {Gonz{\'a}lez-Gait{\'a}n}, S., {Hamuy}, M., {et~al.} 2014,
  \apj, 786, 67

\bibitem[{{Arnett} {et~al.}(2015){Arnett}, {Meakin}, {Viallet}, {Campbell},
  {Lattanzio}, \& {Moc{\'a}k}}]{arnett_mlt_15}
{Arnett}, W.~D., {Meakin}, C., {Viallet}, M., {et~al.} 2015, \apj, 809, 30

\bibitem[{{Bartunov} \& {Blinnikov}(1992)}]{bartunov_blinnikov_92}
{Bartunov}, O.~S. \& {Blinnikov}, S.~I. 1992, Soviet Astronomy Letters, 18, 43

\bibitem[{{Beasor} \& {Davies}(2016)}]{beasor_davies_16}
{Beasor}, E.~R. \& {Davies}, B. 2016, \mnras, 463, 1269

\bibitem[{{Bersten} {et~al.}(2011){Bersten}, {Benvenuto}, \&
  {Hamuy}}]{bersten_11_2p}
{Bersten}, M.~C., {Benvenuto}, O., \& {Hamuy}, M. 2011, \apj, 729, 61

\bibitem[{{Burrows} \& {Goshy}(1993)}]{burrows_goshy_93}
{Burrows}, A. \& {Goshy}, J. 1993, \apjl, 416, L75

\bibitem[{{Chugai} \& {Utrobin}(2000)}]{chugai_utrobin_97D_00}
{Chugai}, N.~N. \& {Utrobin}, V.~P. 2000, \aap, 354, 557

\bibitem[{{Crowther}(2007)}]{pac_wr_07}
{Crowther}, P.~A. 2007, \araa, 45, 177

\bibitem[{{Dall'Ora} {et~al.}(2014){Dall'Ora}, {Botticella}, {Pumo},
  {Zampieri}, {Tomasella}, {Pignata}, {Bayless}, {Pritchard}, {Taubenberger},
  {Kotak}, {Inserra}, {Della Valle}, {Cappellaro}, {Benetti}, {Benitez},
  {Bufano}, {Elias-Rosa}, {Fraser}, {Haislip}, {Harutyunyan}, {Howell},
  {Hsiao}, {Iijima}, {Kankare}, {Kuin}, {Maund}, {Morales-Garoffolo},
  {Morrell}, {Munari}, {Ochner}, {Pastorello}, {Patat}, {Phillips}, {Reichart},
  {Roming}, {Siviero}, {Smartt}, {Sollerman}, {Taddia}, {Valenti}, \&
  {Wright}}]{dallora_12aw_14}
{Dall'Ora}, M., {Botticella}, M.~T., {Pumo}, M.~L., {et~al.} 2014, \apj, 787,
  139

\bibitem[{{Davies} \& {Beasor}(2018)}]{davies_beasor_18}
{Davies}, B. \& {Beasor}, E.~R. 2018, \mnras, 474, 2116

\bibitem[{{Davies} \& {Dessart}(2018)}]{davies_dessart_19}
{Davies}, B. \& {Dessart}, L. 2018, ArXiv e-prints [\eprint[arXiv]{1811.04087}]

\bibitem[{{de Jager} {et~al.}(1988){de Jager}, {Nieuwenhuijzen}, \& {van der
  Hucht}}]{dejager_mdot_88}
{de Jager}, C., {Nieuwenhuijzen}, H., \& {van der Hucht}, K.~A. 1988, \aaps,
  72, 259

\bibitem[{{Dessart} \& {Hillier}(2011{\natexlab{a}})}]{DH11_2p}
{Dessart}, L. \& {Hillier}, D.~J. 2011{\natexlab{a}}, \mnras, 410, 1739

\bibitem[{{Dessart} \& {Hillier}(2011{\natexlab{b}})}]{DH11_pol}
{Dessart}, L. \& {Hillier}, D.~J. 2011{\natexlab{b}}, \mnras, 415, 3497

\bibitem[{{Dessart} {et~al.}(2013){Dessart}, {Hillier}, {Waldman}, \&
  {Livne}}]{d13_sn2p}
{Dessart}, L., {Hillier}, D.~J., {Waldman}, R., \& {Livne}, E. 2013, \mnras,
  433, 1745

\bibitem[{{Dessart} {et~al.}(2018){Dessart}, {Hillier}, \& {Wilk}}]{d18_fcl}
{Dessart}, L., {Hillier}, D.~J., \& {Wilk}, K.~D. 2018, \aap, 619, A30

\bibitem[{{Dessart} {et~al.}(2017){Dessart}, {John Hillier}, \&
  {Audit}}]{d18_13fs}
{Dessart}, L., {John Hillier}, D., \& {Audit}, E. 2017, \aap, 605, A83

\bibitem[{{Dessart} {et~al.}(2010){Dessart}, {Livne}, \& {Waldman}}]{dlw10b}
{Dessart}, L., {Livne}, E., \& {Waldman}, R. 2010, \mnras, 408, 827

\bibitem[{{Falk} \& {Arnett}(1977)}]{FA77}
{Falk}, S.~W. \& {Arnett}, W.~D. 1977, \apjs, 33, 515

\bibitem[{{F{\"o}rster} {et~al.}(2018){F{\"o}rster}, {Moriya}, {Maureira},
  {Anderson}, {Blinnikov}, {Bufano}, {Cabrera-Vives}, {Clocchiatti}, {de
  Jaeger}, {Est{\'e}vez}, {Galbany}, {Gonz{\'a}lez-Gait{\'a}n}, {Gr{\"a}fener},
  {Hamuy}, {Hsiao}, {Huentelemu}, {Huijse}, {Kuncarayakti}, {Mart{\'{\i}}nez},
  {Medina}, {Olivares E.}, {Pignata}, {Razza}, {Reyes}, {San Mart{\'{\i}}n},
  {Smith}, {Vera}, {Vivas}, {de Ugarte Postigo}, {Yoon}, {Ashall}, {Fraser},
  {Gal-Yam}, {Kankare}, {Le Guillou}, {Mazzali}, {Walton}, \&
  {Young}}]{forster_csm_18}
{F{\"o}rster}, F., {Moriya}, T.~J., {Maureira}, J.~C., {et~al.} 2018, Nature
  Astronomy

\bibitem[{{Fraser} {et~al.}(2012){Fraser}, {Maund}, {Smartt}, {Botticella},
  {Dall'Ora}, {Inserra}, {Tomasella}, {Benetti}, {Ciroi}, {Eldridge}, {Ergon},
  {Kotak}, {Mattila}, {Ochner}, {Pastorello}, {Reilly}, {Sollerman},
  {Stephens}, {Taddia}, \& {Valenti}}]{fraser_12aw_12}
{Fraser}, M., {Maund}, J.~R., {Smartt}, S.~J., {et~al.} 2012, \apjl, 759, L13

\bibitem[{{Fraser} {et~al.}(2014){Fraser}, {Maund}, {Smartt}, {Kotak},
  {Lawrence}, {Bruce}, {Valenti}, {Yuan}, {Benetti}, {Chen}, {Gal-Yam},
  {Inserra}, \& {Young}}]{fraser_13ej_14}
{Fraser}, M., {Maund}, J.~R., {Smartt}, S.~J., {et~al.} 2014, \mnras, 439, L56

\bibitem[{{Glas} {et~al.}(2018){Glas}, {Just}, {Janka}, \&
  {Obergaulinger}}]{glas_ccsn_3d_18}
{Glas}, R., {Just}, O., {Janka}, H.-T., \& {Obergaulinger}, M. 2018, ArXiv
  e-prints [\eprint[arXiv]{1809.10146}]

\bibitem[{{Grassberg} {et~al.}(1971){Grassberg}, {Imshennik}, \&
  {Nadyozhin}}]{grassberg_71}
{Grassberg}, E.~K., {Imshennik}, V.~S., \& {Nadyozhin}, D.~K. 1971, \apss, 10,
  28

\bibitem[{{Hamuy}(2003)}]{hamuy_03}
{Hamuy}, M. 2003, \apj, 582, 905

\bibitem[{{Heger} {et~al.}(2000){Heger}, {Langer}, \&
  {Woosley}}]{heger_langer_00}
{Heger}, A., {Langer}, N., \& {Woosley}, S.~E. 2000, \apj, 528, 368

\bibitem[{{Hillier} \& {Dessart}(2012)}]{HD12}
{Hillier}, D.~J. \& {Dessart}, L. 2012, \mnras, 424, 252

\bibitem[{{Hirschi} {et~al.}(2004){Hirschi}, {Meynet}, \&
  {Maeder}}]{hirschi_rot_04}
{Hirschi}, R., {Meynet}, G., \& {Maeder}, A. 2004, \aap, 425, 649

\bibitem[{{Jerkstrand} {et~al.}(2012){Jerkstrand}, {Fransson}, {Maguire},
  {Smartt}, {Ergon}, \& {Spyromilio}}]{jerkstrand_04et_12}
{Jerkstrand}, A., {Fransson}, C., {Maguire}, K., {et~al.} 2012, \aap, 546, A28

\bibitem[{{Jerkstrand} {et~al.}(2014){Jerkstrand}, {Smartt}, {Fraser},
  {Fransson}, {Sollerman}, {Taddia}, \& {Kotak}}]{jerkstrand_12aw_14}
{Jerkstrand}, A., {Smartt}, S.~J., {Fraser}, M., {et~al.} 2014, \mnras, 439,
  3694

\bibitem[{{Kasen} \& {Woosley}(2009)}]{KW09}
{Kasen}, D. \& {Woosley}, S.~E. 2009, \apj, 703, 2205

\bibitem[{{Langer} {et~al.}(1994){Langer}, {Hamann}, {Lennon}, {Najarro},
  {Pauldrach}, \& {Puls}}]{langer_wr_94}
{Langer}, N., {Hamann}, W.-R., {Lennon}, M., {et~al.} 1994, \aap, 290

\bibitem[{{Lentz} {et~al.}(2015){Lentz}, {Bruenn}, {Hix}, {Mezzacappa},
  {Messer}, {Endeve}, {Blondin}, {Harris}, {Marronetti}, \&
  {Yakunin}}]{lentz_ccsn_3d_15}
{Lentz}, E.~J., {Bruenn}, S.~W., {Hix}, W.~R., {et~al.} 2015, \apjl, 807, L31

\bibitem[{{Leonard} {et~al.}(2006){Leonard}, {Filippenko}, {Ganeshalingam},
  {Serduke}, {Li}, {Swift}, {Gal-Yam}, {Foley}, {Fox}, {Park}, {Hoffman}, \&
  {Wong}}]{leonard_04dj_06}
{Leonard}, D.~C., {Filippenko}, A.~V., {Ganeshalingam}, M., {et~al.} 2006,
  \nat, 440, 505

\bibitem[{{Leonard} {et~al.}(2012){Leonard}, {Pignata}, {Dessart}, {Hillier},
  {Horst}, {Fedrow}, \& {Brewer}}]{leonard_12aw_12}
{Leonard}, D.~C., {Pignata}, G., {Dessart}, L., {et~al.} 2012, The Astronomer's
  Telegram, 4033

\bibitem[{{Lisakov} {et~al.}(2017){Lisakov}, {Dessart}, {Hillier}, {Waldman},
  \& {Livne}}]{lisakov_08bk_17}
{Lisakov}, S.~M., {Dessart}, L., {Hillier}, D.~J., {Waldman}, R., \& {Livne},
  E. 2017, \mnras, 466, 34

\bibitem[{{Lisakov} {et~al.}(2018){Lisakov}, {Dessart}, {Hillier}, {Waldman},
  \& {Livne}}]{lisakov_ll2p_18}
{Lisakov}, S.~M., {Dessart}, L., {Hillier}, D.~J., {Waldman}, R., \& {Livne},
  E. 2018, \mnras, 473, 3863

\bibitem[{{Litvinova} \& {Nadezhin}(1985)}]{litvinova_sn2p_85}
{Litvinova}, I.~Y. \& {Nadezhin}, D.~K. 1985, Soviet Astronomy Letters, 11, 145

\bibitem[{{Livne}(1993)}]{livne_93}
{Livne}, E. 1993, \apj, 412, 634

\bibitem[{{Maeder} \& {Meynet}(1987)}]{maeder_meynet_87}
{Maeder}, A. \& {Meynet}, G. 1987, \aap, 182, 243

\bibitem[{{Maguire} {et~al.}(2012){Maguire}, {Jerkstrand}, {Smartt},
  {Fransson}, {Pastorello}, {Benetti}, {Valenti}, {Bufano}, \&
  {Leloudas}}]{maguire_2p_12}
{Maguire}, K., {Jerkstrand}, A., {Smartt}, S.~J., {et~al.} 2012, \mnras, 420,
  3451

\bibitem[{{Maund} {et~al.}(2013){Maund}, {Fraser}, {Smartt}, {Botticella},
  {Barbarino}, {Childress}, {Gal-Yam}, {Inserra}, {Pignata}, {Reichart},
  {Schmidt}, {Sollerman}, {Taddia}, {Tomasella}, {Valenti}, \&
  {Yaron}}]{maund_12ec_13}
{Maund}, J.~R., {Fraser}, M., {Smartt}, S.~J., {et~al.} 2013, \mnras, 431, L102

\bibitem[{{Maund} {et~al.}(2014){Maund}, {Mattila}, {Ramirez-Ruiz}, \&
  {Eldridge}}]{maund_08bk_14}
{Maund}, J.~R., {Mattila}, S., {Ramirez-Ruiz}, E., \& {Eldridge}, J.~J. 2014,
  \mnras, 438, 1577

\bibitem[{{Maund} {et~al.}(2005){Maund}, {Smartt}, \&
  {Danziger}}]{maund_05cs_05}
{Maund}, J.~R., {Smartt}, S.~J., \& {Danziger}, I.~J. 2005, \mnras, 364, L33

\bibitem[{{Meynet} {et~al.}(2015){Meynet}, {Chomienne}, {Ekstr{\"o}m},
  {Georgy}, {Granada}, {Groh}, {Maeder}, {Eggenberger}, {Levesque}, \&
  {Massey}}]{meynet_rsg_15}
{Meynet}, G., {Chomienne}, V., {Ekstr{\"o}m}, S., {et~al.} 2015, \aap, 575, A60

\bibitem[{{Meynet} \& {Maeder}(2000)}]{meynet_maeder_00}
{Meynet}, G. \& {Maeder}, A. 2000, \aap, 361, 101

\bibitem[{{Moriya} {et~al.}(2017){Moriya}, {Yoon}, {Gr{\"a}fener}, \&
  {Blinnikov}}]{moriya_13fs_17}
{Moriya}, T.~J., {Yoon}, S.-C., {Gr{\"a}fener}, G., \& {Blinnikov}, S.~I. 2017,
  \mnras, 469, L108

\bibitem[{{Morozova} {et~al.}(2017){Morozova}, {Piro}, \&
  {Valenti}}]{morozova_2l_2p_17}
{Morozova}, V., {Piro}, A.~L., \& {Valenti}, S. 2017, \apj, 838, 28

\bibitem[{{Morozova} {et~al.}(2018){Morozova}, {Piro}, \&
  {Valenti}}]{morozova_sn2p_18}
{Morozova}, V., {Piro}, A.~L., \& {Valenti}, S. 2018, \apj, 858, 15

\bibitem[{{M{\"u}ller} {et~al.}(2017){M{\"u}ller}, {Melson}, {Heger}, \&
  {Janka}}]{mueller_ccsn_3d_17}
{M{\"u}ller}, B., {Melson}, T., {Heger}, A., \& {Janka}, H.-T. 2017, \mnras,
  472, 491

\bibitem[{{O'Connor} \& {Couch}(2018)}]{oconnor_couch_ccsn_3d_18}
{O'Connor}, E.~P. \& {Couch}, S.~M. 2018, \apj, 865, 81

\bibitem[{{O'Neill} {et~al.}(2019){O'Neill}, {Kotak}, {Fraser}, {Sim},
  {Benetti}, {Smartt}, {Mattila}, {Ashall}, {Callis}, {Elias-Rosa},
  {Gromadzki}, \& {Prentice}}]{oneill_18aoq_19}
{O'Neill}, D., {Kotak}, R., {Fraser}, M., {et~al.} 2019, \aap, 622, L1

\bibitem[{{Paxton} {et~al.}(2011){Paxton}, {Bildsten}, {Dotter}, {Herwig},
  {Lesaffre}, \& {Timmes}}]{mesa1}
{Paxton}, B., {Bildsten}, L., {Dotter}, A., {et~al.} 2011, \apjs, 192, 3

\bibitem[{{Paxton} {et~al.}(2013){Paxton}, {Cantiello}, {Arras}, {Bildsten},
  {Brown}, {Dotter}, {Mankovich}, {Montgomery}, {Stello}, {Timmes}, \&
  {Townsend}}]{mesa2}
{Paxton}, B., {Cantiello}, M., {Arras}, P., {et~al.} 2013, \apjs, 208, 4

\bibitem[{{Paxton} {et~al.}(2015){Paxton}, {Marchant}, {Schwab}, {Bauer},
  {Bildsten}, {Cantiello}, {Dessart}, {Farmer}, {Hu}, {Langer}, {Townsend},
  {Townsley}, \& {Timmes}}]{mesa3}
{Paxton}, B., {Marchant}, P., {Schwab}, J., {et~al.} 2015, \apjs, 220, 15

\bibitem[{{Paxton} {et~al.}(2018){Paxton}, {Schwab}, {Bauer}, {Bildsten},
  {Blinnikov}, {Duffell}, {Farmer}, {Goldberg}, {Marchant}, {Sorokina},
  {Thoul}, {Townsend}, \& {Timmes}}]{mesa4}
{Paxton}, B., {Schwab}, J., {Bauer}, E.~B., {et~al.} 2018, \apjs, 234, 34

\bibitem[{{Popov}(1993)}]{popov_93}
{Popov}, D.~V. 1993, \apj, 414, 712

\bibitem[{{Sanders} {et~al.}(2015){Sanders}, {Soderberg}, {Gezari},
  {Betancourt}, {Chornock}, {Berger}, {Foley}, {Challis}, {Drout}, {Kirshner},
  {Lunnan}, {Marion}, {Margutti}, {McKinnon}, {Milisavljevic}, {Narayan},
  {Rest}, {Kankare}, {Mattila}, {Smartt}, {Huber}, {Burgett}, {Draper},
  {Hodapp}, {Kaiser}, {Kudritzki}, {Magnier}, {Metcalfe}, {Morgan}, {Price},
  {Tonry}, {Wainscoat}, \& {Waters}}]{sanders_sn2_15}
{Sanders}, N.~E., {Soderberg}, A.~M., {Gezari}, S., {et~al.} 2015, \apj, 799,
  208

\bibitem[{{Smartt}(2009)}]{smartt_09}
{Smartt}, S.~J. 2009, \araa, 47, 63

\bibitem[{{Smartt} {et~al.}(2004){Smartt}, {Maund}, {Hendry}, {Tout},
  {Gilmore}, {Mattila}, \& {Benn}}]{smartt_03gd_04}
{Smartt}, S.~J., {Maund}, J.~R., {Hendry}, M.~A., {et~al.} 2004, Science, 303,
  499

\bibitem[{{Sukhbold} {et~al.}(2016){Sukhbold}, {Ertl}, {Woosley}, {Brown}, \&
  {Janka}}]{sukhbold_ccsn_16}
{Sukhbold}, T., {Ertl}, T., {Woosley}, S.~E., {Brown}, J.~M., \& {Janka}, H.-T.
  2016, \apj, 821, 38

\bibitem[{{Turatto} {et~al.}(1998){Turatto}, {Mazzali}, {Young}, {Nomoto},
  {Iwamoto}, {Benetti}, {Cappellaro}, {Danziger}, {de Mello}, {Phillips},
  {Suntzeff}, {Clocchiatti}, {Piemonte}, {Leibundgut}, {Covarrubias}, {Maza},
  \& {Sollerman}}]{turatto_97d_98}
{Turatto}, M., {Mazzali}, P.~A., {Young}, T.~R., {et~al.} 1998, \apjl, 498,
  L129

\bibitem[{{Utrobin}(2007)}]{utrobin_07_99em}
{Utrobin}, V.~P. 2007, \aap, 461, 233

\bibitem[{{Van Dyk} {et~al.}(2012{\natexlab{a}}){Van Dyk}, {Cenko},
  {Poznanski}, {Arcavi}, {Gal-Yam}, {Filippenko}, {Silverio}, {Stockton},
  {Cuillandre}, {Marcy}, {Howard}, \& {Isaacson}}]{vandyk_12aw_12}
{Van Dyk}, S.~D., {Cenko}, S.~B., {Poznanski}, D., {et~al.} 2012{\natexlab{a}},
  \apj, 756, 131

\bibitem[{{Van Dyk} {et~al.}(2012{\natexlab{b}}){Van Dyk}, {Davidge},
  {Elias-Rosa}, {Taubenberger}, {Li}, {Levesque}, {Howerton}, {Pignata},
  {Morrell}, {Hamuy}, \& {Filippenko}}]{vandyk_08bk_12}
{Van Dyk}, S.~D., {Davidge}, T.~J., {Elias-Rosa}, N., {et~al.}
  2012{\natexlab{b}}, \aj, 143, 19

\bibitem[{{Van Dyk} {et~al.}(2003){Van Dyk}, {Li}, \&
  {Filippenko}}]{vandyk_03gd_03}
{Van Dyk}, S.~D., {Li}, W., \& {Filippenko}, A.~V. 2003, \pasp, 115, 1289

\bibitem[{{Vartanyan} {et~al.}(2019){Vartanyan}, {Burrows}, {Radice},
  {Skinner}, \& {Dolence}}]{vartanyan_ccsn_3d_19}
{Vartanyan}, D., {Burrows}, A., {Radice}, D., {Skinner}, M.~A., \& {Dolence},
  J. 2019, \mnras, 482, 351

\bibitem[{{Woosley}(1988)}]{woosley_87A_late_88}
{Woosley}, S.~E. 1988, \apj, 330, 218

\bibitem[{{Woosley} {et~al.}(2002){Woosley}, {Heger}, \& {Weaver}}]{whw02}
{Woosley}, S.~E., {Heger}, A., \& {Weaver}, T.~A. 2002, Reviews of Modern
  Physics, 74, 1015

\bibitem[{{Yaron} {et~al.}(2017){Yaron}, {Perley}, {Gal-Yam}, {Groh}, {Horesh},
  {Ofek}, {Kulkarni}, {Sollerman}, {Fransson}, {Rubin}, {Szabo}, {Sapir},
  {Taddia}, {Cenko}, {Valenti}, {Arcavi}, {Howell}, {Kasliwal}, {Vreeswijk},
  {Khazov}, {Fox}, {Cao}, {Gnat}, {Kelly}, {Nugent}, {Filippenko}, {Laher},
  {Wozniak}, {Lee}, {Rebbapragada}, {Maguire}, {Sullivan}, \&
  {Soumagnac}}]{yaron_13fs_17}
{Yaron}, O., {Perley}, D.~A., {Gal-Yam}, A., {et~al.} 2017, Nature Physics, 13,
  510

\end{thebibliography}

\end{document}